\documentclass[twocolumn,showpacs,preprintnumbers,amsmath,amssymb]{revtex4}
\usepackage[english]{babel}
\usepackage{dcolumn}
\usepackage{amsmath,amssymb,epsfig,float}
\makeatletter

\begin{document}
%\preprint{APS/DP10208}

\title{Nonminimal Maxwell-Chern-Simons-O(3)--$\sigma$ vortices: asymmetric potential case}
\author{M.S. Cunha}
 \email{E-address: marcony@uece.br}
\affiliation {\it Universidade Estadual do Cear\'a, Physics Department, Av.
Paranjana, 1700,%%
60740-000, Fortaleza-Ce, Brazil}%%
\author{R.R. Landim}
\author{C.A.S. Almeida}
 \email{E-address:carlos@fisica.ufc.br}
\affiliation {%
Universidade Federal do Cear\'{a}, Physics Department, C.P. 6030, 60455-760
Fortaleza-Ce, Brazil} %%%

\begin{abstract} In this work we study a nonlinear gauged O(3)-sigma model
with both minimal and nonminimal coupling in the covariant derivative. Using an asymmetric scalar
potential, the model is found to exhibit both topological and non-topological soliton solutions in
the Bogomol'nyi limit.%
\end{abstract}

% PACS, the Physics and Astronomy
                             % Classification Scheme.
%\keywords{Suggested keywords}%Use showkeys class option if keywo
                              %display desired
\maketitle
\section{The (2+1) nonminimal model}

The $O(N)- \sigma$ model is a generalization of the $O(4)$ model, introduced by Gell-Mann and Levy
\cite{gell-mann}. Classically, the field configurations are $\varphi$ mappings from a space-time
to a target space, which is a multiplet of N scalar real fields $\varphi _{i},$ $i=1,...,N,$ under
the constraint $\varphi ^{2}=\varphi _{i}\varphi _{i}=1$.

Since solitonic solutions have some importance in Condensed Matter systems
\cite{polyakov,laughlin,zee}, especially in two-dimensional isotropic ferromagnets
\cite{belavin,Rajaraman}, some authors have studied gauge invariance of the $O(3)$-model where the
group $U(1)$ is a subgroup of $O(3)$.

However, since these studies, Stern has shown that the inclusion of nonminimal coupling in
Maxwell-Chern-Simons electrodynamics tends to mimic anyonic behavior without the pure Chern-Simons
limit \cite{stern}.

In the $O(3)-\sigma$ model, the complex scalar field $\pmb \phi$ is a 3-vector constrained to
satisfy the relation $\pmb \phi \cdot \pmb {\phi}=\phi _{1}^{2} +\phi _{2}^{2}+\phi _{3}^{2}=1.$

We are looking for a Lagrangian invariant under global iso-rotations of the field $\pmb \phi$
about a fixed axis $\pmb {n}$ $\in S^{2}.$ In order to gauge the symmetry, we choose $\pmb
{n}=(0,0,1)$. The 3D-gauge invariant Lagrangian density is then given by
\begin{eqnarray}
\mathcal{L} =  -\frac{1}{4}F_{\mu \nu }^{2}+\frac{\kappa }{4}\varepsilon^{\mu \nu \rho}A_{\mu
}F_{\nu \rho}+\frac{1}{2}\nabla _{\mu }\pmb {\phi }\cdot \nabla ^{\mu }\pmb {\phi }~~~~~~~~~~
\nonumber\\
+\frac{1}{2}\partial _{\mu }M\partial ^{\mu }M - \frac{1}{2}g\partial _{\mu}M\partial^{\mu }(\pmb
{n\cdot \phi })-U(M, \pmb {n}\cdot \pmb {\phi})\label{one}
\end{eqnarray}

The first two terms in Eq. (\ref{one}) are the normal Maxwell and Chern-Simons terms, $M$ is a
neutral scalar field and $\pmb \phi$ is the multiplet field, as already defined. The introduction
of the $M$ field \cite{LLM} can be explained by supersymmetry arguments~\cite{marc}. The gauge
covariant derivative is modified in order to introduce non-minimal coupling, \textit{i.e.},
\begin{equation}
\nabla _{\mu }\pmb {\phi }=\partial _{\mu }\pmb {\phi }+\left[eA_{\mu }+%%
\frac{g}{2}\varepsilon _{\mu \nu \alpha }F^{\nu \alpha }\right]\pmb {n}\times
\pmb {\phi }\label{two}\end{equation}

The real parameter $g$ introduced in Eq. (\ref{one}) and in the covariant derivative is a
nonminimal coupling constant. The equation of motion for the gauge field $A_{\mu}$ can be written
as

\begin{equation}
\partial _{\mu }\left[\varepsilon ^{\mu \alpha \nu }\left(\frac{g}{e}J_{\nu
}+%%
F_{\nu }\right)\right]=J^{\alpha }-\kappa F^{\alpha
},\label{five}\end{equation} %
where we have used the $F^{\mu }=\frac{1}{2}\varepsilon^{\mu \nu \alpha }F_{\nu
\alpha }$ dual
field and the matter current %%
 $J^{\mu }=-e\, \pmb {n\cdot \phi }\times \nabla ^{\mu }\pmb {\phi }.$
Assuming a critical coupling $g_{c}=-\frac{e}{\kappa },$ we obtain a
topological first order
equation \cite{JLWPRL,JLW},
\begin{equation}
F^{\alpha }=\frac{1}{\kappa }J^{\alpha }\; .\label{six}\end{equation}%%%
The above equation was first used by Stern \cite{stern} and later by Torres \cite{torres,
escalona}.

As in \cite{kimm}, we can now construct a general topological gauge invariant current
\begin{equation}%%
J_{top}^{\mu }=\frac{1}{8\pi }\varepsilon ^{\mu \nu \alpha }\pmb {\phi }\cdot
\left[D_{\nu }%%
\pmb {\phi }\times D_{\alpha }\pmb {\phi }-\frac{e}{2}F_{\nu \alpha
}\left(v-\pmb {n}\cdot %%
\pmb {\phi }\right)\pmb {\phi }\right]\label{topcurr}\end{equation} with $v$ as a real parameter.
This current yields a topologically conserved charge given by
\begin{equation} Q_{top}=\frac{1}{4\pi }\int
d^{2}x\left[\pmb {\phi }\cdot D_{1}\pmb %%
{\phi }\times D_{2}\pmb {\phi }+\frac{eB}{2}\left(v-\pmb {n\cdot \phi
}\right)\right].\label{topcharg}%%
\end{equation}%%%%

We can impose different boundary conditions at spatial infinity for finite energy field
configurations. Depending on such conditions, the solutions belong to a broken phase or to a
symmetric one. In the $v=1$ symmetric phase we have $\lim _{|\pmb {x}|\rightarrow \infty }\pmb
{\phi }(t,\pmb {x})=\pm \pmb {n}\label{lim1}$. On the other hand, for $\left|v\right|<1,$ we can
impose $\lim _{|\pmb {x}|\rightarrow \infty }\pmb {n\cdot }\pmb {\phi }(t,\pmb {x})=v$ which
characterizes the asymmetric (broken) phase.

\section{Bogomol'nyi equations}

The derivative of the Lagrangian (\ref{one}) with respect to the space-time metric tensor $g_{\mu
\nu }$ produces a naturally symmetric tensor in the $\mu $ and $\nu$ indices, namely,
\begin{eqnarray} %%
&T_{\mu \nu}= G(g,\pmb {\phi }{\textbf {)}}F_{\mu }^{\; \alpha }F_{\alpha \nu
}-\frac{1}{2}\left[\partial _{\mu }M\partial _{\nu }(\pmb {n\cdot \phi })%%
 +\partial _{\nu }M\partial _{\mu }(\pmb {n\cdot \phi })\right]&\nonumber%%
 \\
 &+D_{\mu }\pmb {\phi }\cdot D_{\nu }\pmb {\phi }+\partial _{\mu }M\partial
_{\nu }M %%
 -\eta _{\mu \nu }\mathcal{L}_{ntop}\mbox %%
 {\;,}~~~~~~~~~~~~~~&\label{emt}
\end{eqnarray}%%%
where $G(g,\pmb {\phi }{\textbf {)}}=1-g^{2}\left(\pmb {n}\times \pmb {\phi }\right)^{2}$ and
$\eta_{\mu \nu}$ stands for flat metric $(1,-1,-1)$. $\mathcal{L}_{ntop}$ differs from
$\mathcal{L}$ only in the topological terms. The energy functional is given by
\begin{eqnarray}
&T_{00}=\frac{1}{2}G\left(B^{2}+E^{2}\right)+\frac{1}{2}\partial _{0}M\partial
_{0}M+%%
\frac{1}{2}\partial _{i}M\partial _{i}M\nonumber \\
&-g\partial _{0}M\partial _{0}(\pmb {n\cdot \phi })-%
 g\partial _{i}M\partial _{i}(\pmb {n\cdot \phi })+\frac{1}{2}D_{0}\pmb {\phi
}%%
 \cdot D_{0}\pmb {\phi }&\nonumber\\
&+\frac{1}{2}D_{i}\pmb {\phi }\cdot D_{i}\pmb {\phi }+U(M,\pmb{n\cdot \phi
})\mbox
{.}~~~~~~~~~~~~~~~~~~~~~~~~~& \label{emt0}
\end{eqnarray}

We can establish the following expression
\begin{eqnarray}
&\frac12 D_{i}\pmb {\phi }\cdot D_{i}\pmb {\phi } =\frac{-g^{2}}2 \left(\pmb
{n}\times%%
\pmb {\phi }\right)^{2}E^{2}+\frac{g}{e}F_{i}k_{i}%%
\mp gE_{i}\partial _{i} (\pmb {n\cdot \phi })&\nonumber \\
&~~~~~~~~~~~+\frac{1}{2} (\nabla _{1}\pmb {\phi }\,{\pm }\,\pmb {\phi } \times \nabla_{2}\pmb{\phi
})^{2}\pm \pmb {\phi }\cdot D_{1}\pmb {\phi }\times D_{2}\pmb {\phi } \mbox {,}&\label{emt10}
\end{eqnarray}
where we have used $k^{\mu }\equiv -e\pmb {n\cdot \phi }\times $ $D^{\mu }\pmb {\phi }$, which
contains no explicit contribution of the nonminimal coupling. Using Eq.~(\ref{emt10}) in
Eq.~(\ref{emt0}) and integrating it over the whole space, we obtain the total energy
$\mathcal{E}$,
 \begin{eqnarray} %%
 \mathcal{E} = \int d^{2}x\left\{\frac{1}{2}G\left[B\mp
\frac{e}{G}\left(v-\pmb {n\cdot %%
\phi }+\frac{\kappa }{e}M\right.\right.\right.~~~~~~~~~~~~~~~&&\nonumber\\
+\left. \left. \left.g\left(\pmb {n}\times \pmb {\phi
}\right)^{2}M\right)\right]^{2}\right.%%
\pm eB\left[v-\pmb {n\cdot \phi }+\frac{\kappa
}{e}M\right.~~~~~~~~~~~~~&\nonumber\\
\left.+g\left(\pmb {n}\times \pmb {\phi %%
 }\right)^{2}M\right]+\frac{1}{2}\left(E_{i}\pm \partial _{i}M\right)^{2}\mp
E_{i}\partial
_{i}M~~~~~~~~~~~~~~&\nonumber \\%%
-\frac{1}{2}g^{2}\left(\pmb {n}\times \pmb {\phi }\right)^{2}%%
 E^{2}+\frac{1}{2}\partial _{0}M\partial _{0}M-\frac{1}{2}g\partial
_{0}M\partial _{0}(\pmb %%
 {n\cdot \phi })~~~~&\nonumber \\
 -g\partial _{i}M\partial _{i}(\pmb {n\cdot \phi })+\frac{1}{2}|D_{0}\phi
 \pm eM\left(\pmb {n}\times \pmb {\phi }\right)|^{2}~~~~~~~~~~~~~~~~\nonumber
\\
 \pm M\left(J_{0}-egB\left(\pmb {n}\times \pmb {\phi
}\right)^{2}\right)-\frac{1}{2}g^{2}
 \left(\pmb {n}\times \pmb {\phi }\right)^{2}E^{2}~~~~~~~~~~~~&\nonumber \\
 +\frac{g}{e}F_{i}k_{i}\mp gE_{i}\partial _{i}\left(\pmb {n\cdot \phi }\right)
 +\frac{1}{2}\left(\nabla _{1}\pmb {\phi } %%
 {\pm }\pmb {\phi }\times \nabla _{2}\pmb {\phi }\right)~~~~~~~~~~~~&\nonumber
\\
 \left.\pm \pmb {\phi }\cdot D_{1}%%
 \pmb {\phi }\times D_{2}\pmb %%
 {\phi }+U(M,\pmb{n\cdot \phi })\right\} \mbox
{.}~~~~~~~~~~~~~~~~~~~~~~~~~~~&\label{energy}
\end{eqnarray}%%%
%%%%%%%%%%%%%%%%%%%%%%%%%%%%%%%%%%%%%%%%%
In order to achieve the Bogomol'nyi limit, we must choose an adequate potential
given by
\begin{eqnarray}
U&=&\frac{e^{2}}{2G}\left[v-\pmb {n\cdot \phi }+\frac{\kappa }{e}M+g\left(\pmb {n}\times %%
\pmb {\phi }\right)^{2}M\right]^{2} \nonumber\\%%
&&+\frac{1}{2}e^{2}M^{2}\left(\pmb {n}\times \pmb{\phi } \right)^{2}\label{sixteen}
\end{eqnarray}

The above potential is a generalization of potentials found in the literature, involving Maxwell
and Chern-Simons terms \cite{kimm,ghosh1,muk,muk2}. In the present case, the following equations
are necessary to achieve the energy bound limit,
\begin{subequations}
\begin{eqnarray} %
B\mp \frac{e}{G}\left[v-\pmb {n\cdot \phi }\frac{\kappa }{e}M+g\left(\pmb {n}\times \pmb %%
{\phi }\right)^{2}M\right] & = & 0\label{sd1}\\
E_{i}\pm \partial _{i}M & = & 0\label{sd2}\\
D_{0}\pmb {\phi }\pm eM\left(\pmb {n}\times \pmb {\phi }\right) & = &
0\label{sd3}\\
\nabla _{1}\pmb {\phi }\pm \pmb {\phi }\times \nabla _{2}\pmb %%
{\phi } & = & 0\mbox {.}\label{sd4}
\end{eqnarray}
\end{subequations}%
These are the Bogomol'nyi (or selfdual) equations \cite{bogo} for the MCS-O(3)-$\sigma$ model with
nonminimal coupling in the asymmetric potential case. As usual, if we consider static solutions,
the total energy is simplified, namely,
\begin{equation}
\mathcal{E} = 4\pi
\left|Q_{top}\right|{.}\label{bound}
\end{equation}

\section{Critical case $g=g_{c}$}

The selfdual Eqs. (\ref{sd1})-(\ref{sd4}) are strongly coupled and not easily solvable. However
they are simplified by considering the critical limit $g=g_{c}.$ In this case, $M$ and $B$
fields decouple and can be written as
\begin{subequations}%
\begin{eqnarray} %%
&M=g_{c}\left(v-\pmb {n}\cdot \pmb {\phi}\right).&\\
&B=\pm eg_{c}^{2}\frac{(\pmb {n}\times \pmb
{\phi })^{2}}{1-g_{c}^{2}(\pmb {n}\times %%
\pmb {\phi })^{2}}\left(v-\pmb {n}\cdot \pmb {\phi }\right)&\label{B0}
\end{eqnarray}%%%
\end{subequations}%%%
As a consequence, the potential (\ref{sixteen}) can be written only in terms of $\pmb {\phi} $ as
\begin{equation}
U=\frac{e^{2}g_{c}^{2}}{2}\frac{(v-\pmb {n}\cdot \pmb {\phi
})^{2}}{1-g_{c}^{2}%%
(\pmb {n}\times \pmb {\phi })^{2}}\left(\pmb {n}\times \pmb {\phi }\right)^{2},%%
\label{critpot}\end{equation} %
as illustrated in Fig. (\ref{fig1}) for $g=g_c=-0.5$. For arbitrary $v$,
when we substitute $g_{c}=-e/\kappa $ and then take $\kappa
>>1$, the above expression reduces to the case studied in Ref.~\cite{kimm},
\begin{equation}
U \approx U_{CS}=\frac{e^{4}}{2\kappa ^{2}}(v-\pmb {n}\cdot \pmb {\phi
})^{2}\left(\pmb %%
{n}\times \pmb {\phi }\right)^{2}.\label{Ucs}\end{equation} %%%%%%%%%%%%
%%%%%%%%%%%%%%%%%%%%%%figure%%%%%%%%%%%%%%%%%%%%%%
 \unitlength=1cm %
\begin{figure}
\centering
\begin{picture}(7,5)
\epsfig{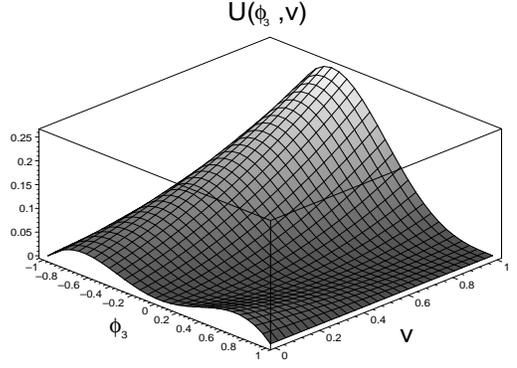}
\end{picture}
\caption{\protect{\small The three dimensional scalar potential $U(\phi _{3},v)$ with
$g_{c}=-0.5$. Note that the potential is symmetric in $v=0$.}} \label{fig1}
\end{figure}
%%%%%%%%%%%%%%%%%%%%%%%%%%%%%%%%%%%%%%%%%%%%%%%%%%%%
\section{Static vortices in the $g=g_{c}$ limit}
In order to solve the remaining Eqs. (\ref{sd4}) and (\ref{B0}), we use a well-known
parameterization for $\pmb {\phi }$, which is valid for invariant solutions under both rotations
and reflections in space-time and target space manifold \cite{gladi1}. Explicitly,
\begin{equation}
\pmb {\phi }(r,\theta )=\left(\sin f\cos N\theta ,\; \sin f\sin N\theta ,\; %%
\cos f\right)\label{ansatz1}\end{equation} %%%%%%%%
where $f=f(r)$ and $N$ is an integer which defines the topological grade (vorticity) of the
solution. For the gauge field we also use a parameterization
which takes into account the symmetries of the model, \textit{i.e.,}%%%%%
%%%%%%%%%%%%%
\begin{eqnarray} %%%
{\bf A}=\frac{a(r)-N}{er}\hat{\mathbb\theta }\; .\label{ansatz2}%%%
\end{eqnarray} %
%%%%%%%%%%%%%%%
With this choice, we obtain two coupled nonlinear first order differential
equations,
\begin{subequations}
\begin{eqnarray} %
&f^{\prime }(r)=\pm \frac{1}{r}\frac{\sin f(r)}{1+g_{c}^{2}\sin
^{2}f(r)}a(r)&%%
\label{f(r)}\\ %
&a^{\prime }(r)=\pm r\frac{g_{c}^{2}\sin ^{2}f(r)}{1-g_{c}^{2}\sin ^{2}f(r)}\;
\left[v-\cos f(r)\right]\mbox {,}&\label{a(r)}%
\end{eqnarray} \label{dif}%%%%%
\end{subequations}%%%%%%%%
where we have used $A_{0}=\mp M$ and $r\rightarrow \frac{1}{e}r$.

\section{Analysis and Numerical results}

Equations (\ref{f(r)}) and (\ref{a(r)}) have no analytical solution. However, we can analyze their
asymptotic behavior. Considering Eq. (\ref{ansatz2}), non singular solutions at the origin always
require that, for small values of $r$, $a(r)$ approaches $N$. Assuming $f(r\rightarrow0)<<1$ then,
from Eqs.~(\ref{dif}), $f(r)\simeq C_N r^N$ and $a(r)\simeq
N+(v-1)C^2_Ng^2_{crit}r^{2N+2}/(2N+2)$, $N \in \mathbb{Z}$*.

When $r\rightarrow \infty$, $a^{\prime }(r)$ cannot diverge, and thus we have three situations
satisfying Eqs.~(\ref{dif}), namely $ f(\infty )= (0, \pi, f_v)$, where we follow Ref.~\cite{kimm}
in labelling $f(\infty )= f_{v}$ such that $cosf_v=v$. These values correspond to the minima of
the potential (\ref{critpot}) in the interval {[}-1,1{]}.

The $C_{_{N}}\in \mathbb{R}$ constants are determined by the behavior of $f(r)$ at spatial
infinity. Our numerical results indicate that there exists a critical value for $C_{N}$ such that
$f(\infty )\rightarrow 0$ ($C_{_{N}}<C_{crit} $); $f(\infty)\rightarrow \pi$
($C_{_{N}}>C_{crit}$); and $f(\infty )\rightarrow f_{v}$ ($C_{_{N}}=C_{crit}$).

Soliton solutions are similar to those of Ref.~\cite{kimm}. When
 $C_N=C_{crit}$, the ($N\neq0$) vortices are in the asymmetric phase of the potential and the
magnetic flux given by $\Phi =\frac{2\pi }{e}N$, becomes quantized.

%\unitlength=1cm
\begin{figure}[hb]
\centering
\begin{picture}(6,6)
\epsfig{file=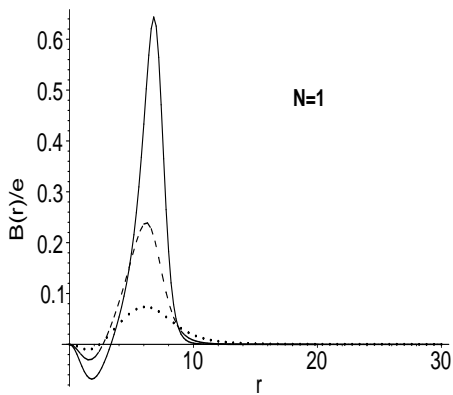,width=6cm,height=6cm}%%
%%%\put(-1.7,0.55){\small{$v$}} \put(-5.3,0.4){$^{\phi_3}$}
\end{picture}
\caption{{\protect\small The magnetic field $B(r)$ for a few values of $g_{c}$ ($v=0.5$) when
$f(\infty )=\protect\pi $ (symmetric phase of the potential), for $N=1$ vorticity. The solid line
represents $B(r)$ for $g_{c}=-0.7$, dashed line $g_{c}=-0.5$, and dotted line $g_{c}=-0.3$. }}
\label{fig2}%%%
\end{figure}
%%%%end figure
When $C_{_{N}}\neq C_{crit}$, topological (and nontopological) vortices are in the symmetric phase
of the potential and the magnetic flux, now given by $\Phi =\frac{2\pi }{e}\left(N-\alpha
\right)$, varies continuously, where $\alpha $ is the asymptotic value of $a(r)$ at spatial
infinity. Numerical analysis shows that for $C_{_{N}}<C_{crit},$ one has $\alpha <0$; for
$C_{_{N}}>C_{crit}$, one has $\alpha > 0$.%%
%%%%%%%%%%%

%%%%%%%%%%%%%%
%%%%%%%%figure
Fig. (\ref{fig2}) shows the magnetic field for $C_N>C_{crit}$ case, for a few values of $g_c$ with
$N=1$. Other topological solutions ($|N|>1$) exhibit similar behavior. Note that the magnetic
field changes sign.
%It is a topological lump in the symmetric phase $f(\infty)=\pi$.
%%%%%%%%%%%%%%%%%%%%%%%%%%%%%%figure%%%%%%%%%%%%%%%%%%
\unitlength=1cm
\begin{figure}[h]
\begin{picture}(7,6) \epsfig{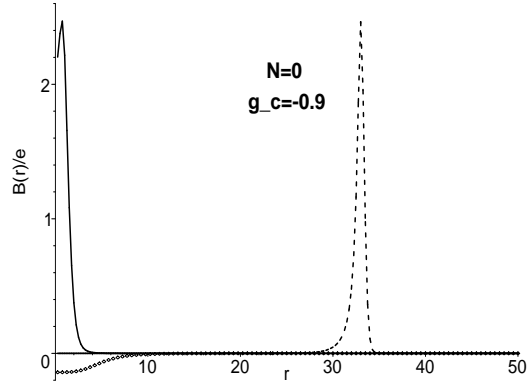}%%
%%%\put(-1.7,0.55){\small{$v$}} \put(-5.3,0.4){$^{\phi_3}$}
\end{picture}
\caption{{\protect\small The magnetic field $B(r)$ for null vorticity for different values of
$f(r)$ at the origin (v=0.5): $f(0)=\pi/4$ ($\diamond$) with $B(r)$'s maximal value at the origin;
$f(0)=\pi/3$ (dashed line); $f(0)=\pi/2$ (solid line). }}%%
\label{fig3}
\end{figure}
%%%%%%%%%%%%%
\begin{figure}[h]
\centering
\begin{picture}(7,6)
\epsfig{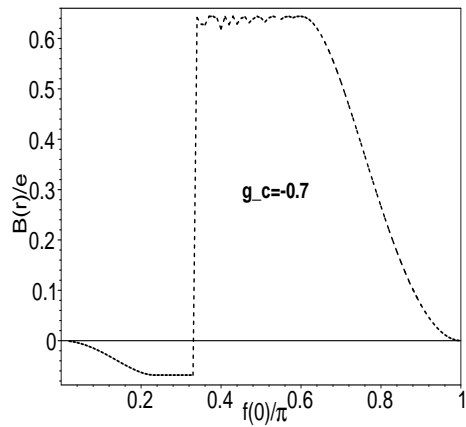}
\end{picture}
{\vskip0.3cm } \caption{\protect{\small Maxima values of the magnetic field $B(r)$} for different
values of {\small $f(0)/\pi $ when $g_{c}=-0.7$ ($v=0.5$) and null voticity (N=0). Note an abrupt
change in $B(r)$ at $f(0)=\pi /3$.}} \label{fig4}
\end{figure}
%%%%%%%%%%%%%%%%%%%%%%%%%
%%%%%%%%%%%%%%%%%%%%%%%%%%%%%%figure%%%%%%%%%%%%%%%%%%

Solutions with null vorticity $\left(N=0\right)$ have neither quantized topological charge nor
quantized flux but they are of some interest. The magnetic field shows some interesting
characteristics for the null vorticity. As seen in Fig. (\ref{fig3}), $|B(r)|$ has a maximum at
the origin for $f(0) < \pi/3$. On the other hand, at $f(0)=\pi/3$ the maximum appears away
from the origin, returning as $f(0)$ increases. In general, the values of the maxima depend on
$g_{c}$ and the conditions on $f(r)$ at the origin {[}Fig. (\ref{fig4}){]}.
%%%%%%%%
\section{Conclusions}
To summarize, in this work we construct a gauged non-linear-O(3)-$\sigma$ model. The gauging of
the U(1) subgroup includes nonminimal coupling in the covariant derivative. The gauge-field
dynamics are governed by both Maxwell and Chern-Simons terms.

Through the Bogomol'nyi method, we find an asymmetric potential that contains the contribution of
the nonminimal term, and the self-dual equations necessary to achieve the energy bound limit.

It is worthwhile to note that when one chooses a specific value for the nonminimal coupling
constant, namely, $g=-\frac e\kappa$, some fields decouple and the Higgs potential changes,
leading the system to behave like a pure Chern-Simons system. However the model proposed here is
different since it possesses both Maxwell and Chern-Simons terms.

The solitons found have topological and non-topological characteristics. When we compare our
results with those of models without nonminimal coupling, we can see notable differences
in the magnetic flux intensity. This implies that all the fields are influenced by the
value of the nonminimal coupling constant.

Indeed, the magnetic field shows interesting behavior in our model. The nontopological solutions
in the symmetric case show a magnetic field maximum at the origin and moved away from it when
$f(0)\geq\pi/3$. Topological solutions in symmetric phase show that magnetic flux is not quantized
and magnetic field changes sign. This behavior is related to topological lumps \cite{kimm}.

Topological vortices were found in the asymmetric phase where $\pmb {n\cdot }\pmb {\phi
}(\infty)\rightarrow v$. In this case, the magnetic field is purely attractive or repulsive and
the flux is quantized.

A further investigation could also be performed for two dimensional models obtained from a
dimensional reduction. These kinds of reduced models can exhibit interesting domain wall
structures.

\begin{acknowledgments}
The authors would like to thank Funda\c{c}\~ao Cearense de Apoio ao Desenvolvimento Cient\'ifico e
Tecnol\'ogico do Estado do Cear\'a - FUNCAP, and CNPq (Brazilian Research Agency) for financial
support, and Dr. H.R. Christiansen and Dr. A. Donegan for a critical reading of the manuscript.
MSC dedicates this work to ATLR, who was a very special person in his life.
\end{acknowledgments}

\end{document}